\documentclass[a4paper]{PoS}

\usepackage[numbers]{natbib}
\usepackage{rotating}
\usepackage{graphicx}
\usepackage{fixltx2e}
\usepackage{subfig}

\title{Very high energy emission from the hard spectrum sources HESS\,J1641$-$463, HESS\,J1741$-$302 and HESS\,J1826$-$130}

\ShortTitle{VHE emission from HESS\,J1641$-$463, HESS\,J1741$-$302 and HESS\,J1826$-$130}

\author{\speaker{E. O. Ang\"uner}$^{a}$, S. Casanova$^{a,b}$, I. Oya$^{c}$, F. Aharonian$^{b,d}$, P. Bordas$^{b}$ and A. Ziegler$^{e}$ for the H.E.S.S. Collaboration.\\
\llap{$^a$}Instytut Fizyki J\c{a}drowej PAN, ul. Radzikowskiego 152, 31-342 Krak{\'o}w, Poland\\
\llap{$^b$}Max-Planck-Institute fr Kernphysik, P.O. Box 103980, D 69029 Heidelberg, Germany\\
\llap{$^c$}DESY, Platanenallee 6, D-15738 Zeuthen, Germany \\
\llap{$^d$}Dublin Institute for Advanced Studies, 31 Fitzwilliam Place, Dublin 2, Ireland \\
\llap{$^e$}Friedrich-Alexander-Universit\"at Erlangen-N\"urnberg, Erlangen Centre for Astroparticle Physics, Erwin-Rommel-Str. 1, D 91058 Erlangen, Germany \\

E-mail: \email{oguzhan.anguner@ifj.edu.pl}, \\
\email{sabrina.casanova@mpi-hd.mpg.de}, \\
\email{igor.oya.vallejo@desy.de}, \\
\email{felix.aharonian@mpi-hd.mpg.de}, \\
\email{pol.bordas@mpi-hd.mpg.de},\\
\email{alexander.ziegler@fau.de}}

\abstract{A recent study of the diffuse $\gamma$-ray emission in the Central Molecular Zone using very high energy (VHE, E $>$ 0.1 TeV) H.E.S.S. data suggests that the 
Galactic Center (GC) is the most plausible supplier of Galactic ultra-relativistic cosmic-rays (CRs) up to the knee at about 10$^{15}$ eV (PeV). However, the GC might not 
be the only source capable to accelerate CRs up to PeV energies in the Galaxy. Here we present H.E.S.S. data analysis results and interpretation of three H.E.S.S. sources, 
with spectra extending beyond 10 TeV and relatively hard spectral indices compared with the average spectral index of H.E.S.S. sources, namely HESS\,J1641$-$463, HESS\,J1741$-$302 
and HESS\,J1826$-$130. Although the nature of these VHE $\gamma$-ray sources is still open, their spectra suggest that the astrophysical objects producing such emission must 
be capable of accelerating the parental particle population up to energies of at least several hundreds of TeV. Assuming a hadronic scenario, dense gas regions can provide 
rich target material for accelerated particles to produce VHE $\gamma$-ray emission via proton-proton interactions followed by a subsequent $\pi^{0}$ decay. Thus, detailed 
investigations of the interstellar medium along the line of sight to all of these sources have been performed by using data from available atomic and molecular hydrogen 
surveys. The results point out the existence of dense interstellar gas structures coincident with the best fit positions of these sources. One can find possible hadronic 
models with CRs being accelerated close to the PeV energies to explain the $\gamma$-ray emission from all of these sources, which opens up the possibility that a population 
of PeV CR accelerators might be active in the Galaxy.}

\FullConference{35th International Cosmic Ray Conference --- ICRC2017\\
		10--20 July, 2017\\
		Bexco, Busan, Korea}

\begin{document}

\section{INTRODUCTION}


Investigation of the diffuse very high energy (VHE, E $>$ 0.1 TeV) H.E.S.S. $\gamma$-ray data around the Galactic Centre (GC) region suggests the presence of PeV 
(${10}^{15}$ eV) particles within the central 10 parsec of the Galaxy, making the GC the most plausible supplier of ultra-relativistic cosmic-rays (CRs), a 
``PeVatron'' \citep{gcpev}. However, there may be other sources contributing to the Galactic CR flux around the knee, representing a new source population. One expects that 
such sources, capable of accelerating protons (hadrons) up PeV energies, should have relatively hard spectral indices without showing any cut-off in their VHE $\gamma$-ray 
spectra. The H.E.S.S. Galactic Plane Scan (HGPS) catalog \citep{hgps} can be used to pin-point sources with spectra extending beyond at least 10 TeV. Firm identification of 
the nature of such sources is quite important since they can provide information to understand a long-standing question in science, the origin of the highest energy CRs up 
to PeV energies.

In this work, we present dedicated VHE and multi-wavelength (MWL) data analysis results and interpretation of three intriguing H.E.S.S. sources, HESS\,J1641$-$463 \citep{igor}, 
HESS\,J1741$-$302 \citep{tibolla} and HESS\,J1826$-$130 \citep{oziJ1826}. These three H.E.S.S. sources show common characteristics, especially from the spectral point of view, 
as listed below.   

\begin{itemize}
\item Inspection of their VHE $\gamma$-ray spectra suggests that the astrophysical objects producing such emission must be capable of accelerating the parental particle 
population up to energies of at least several hundreds of TeV.
\item All of them are tagged as unidentified sources and located near other bright VHE sources, thus their (apparent) emission is contaminated and suffering from source 
confusion. 
\item Their best fit positions and extensions are spatially coincident with dense gas regions.
\item None of the sources shows variable VHE emission. 
\end{itemize}

\section{H.E.S.S. OBSERVATIONS and RESULTS}

\subsection{The H.E.S.S. Telescopes}

The High Energy Stereoscopic System (H.E.S.S.) is an array of five imaging atmospheric Cherenkov telescopes located in the Khomas Highland of Namibia, 1800 m above sea level. H.E.S.S. in phase I comprised of 
four 13 m diameter telescopes which have been fully operational since 2004. A fifth telescope was added in the center of the array and has been operational since September 
2012. The H.E.S.S. phase I array configuration is sensitive to $\gamma$-ray energies between 100 GeV and several tens of TeV. With the addition of the fifth telescope, the 
energy threshold was lowered down to some tens of GeV. The VHE H.E.S.S. data presented in this paper were taken with the H.E.S.S. phase I array configuration, which can measure 
extensive air showers with an angular resolution better than ${0.1}^{\circ}$ and an energy resolution of 15$\%$ at an energy of 1 TeV \citep{Aharonian06}.

\subsection{H.E.S.S. Analysis Details}

The data presented here have been analyzed with the H.E.S.S. analysis package for shower reconstruction. For the case of HESS\,J1741$-$302 and HESS\,J1826$-$130, the 
multivariate analysis (TMVA) technique \citep{TMVA} has been applied for providing the best available rejection power between hadrons and $\gamma$ rays, while this technique has 
not been applied for the case of HESS\,J1641$-$463. The CR background levels were estimated using the ring background model \citep{berge2007} for source detection and 
morphology studies. The detection significances were determined by using Equation (17) in \cite{lima}. The differential VHE $\gamma$-ray spectra of these sources were 
derived using the forward folding technique \citep{piron2001} and the reflected background model \citep{berge2007} was used for the background estimation. A log-likelihood 
ratio test (LLRT) was used for comparing both the morphology and spectral models and deciding the preferred models that describe the data best.

\subsection{Morphology Results}

Thanks to the excellent sensitivity of H.E.S.S., combined with the high density of sources in the Galaxy, the number of detected sources increases significantly. But this 
leads to confusion of sources, especially for the ones located in the vicinity of bright sources. All three sources presented here suffer from source confusion in different 
levels, which means that they are contaminated by the emission coming from nearby bright sources. 

\begin{figure}[ht!]
\centering
\subfloat[]{{\includegraphics[width=4.4cm]{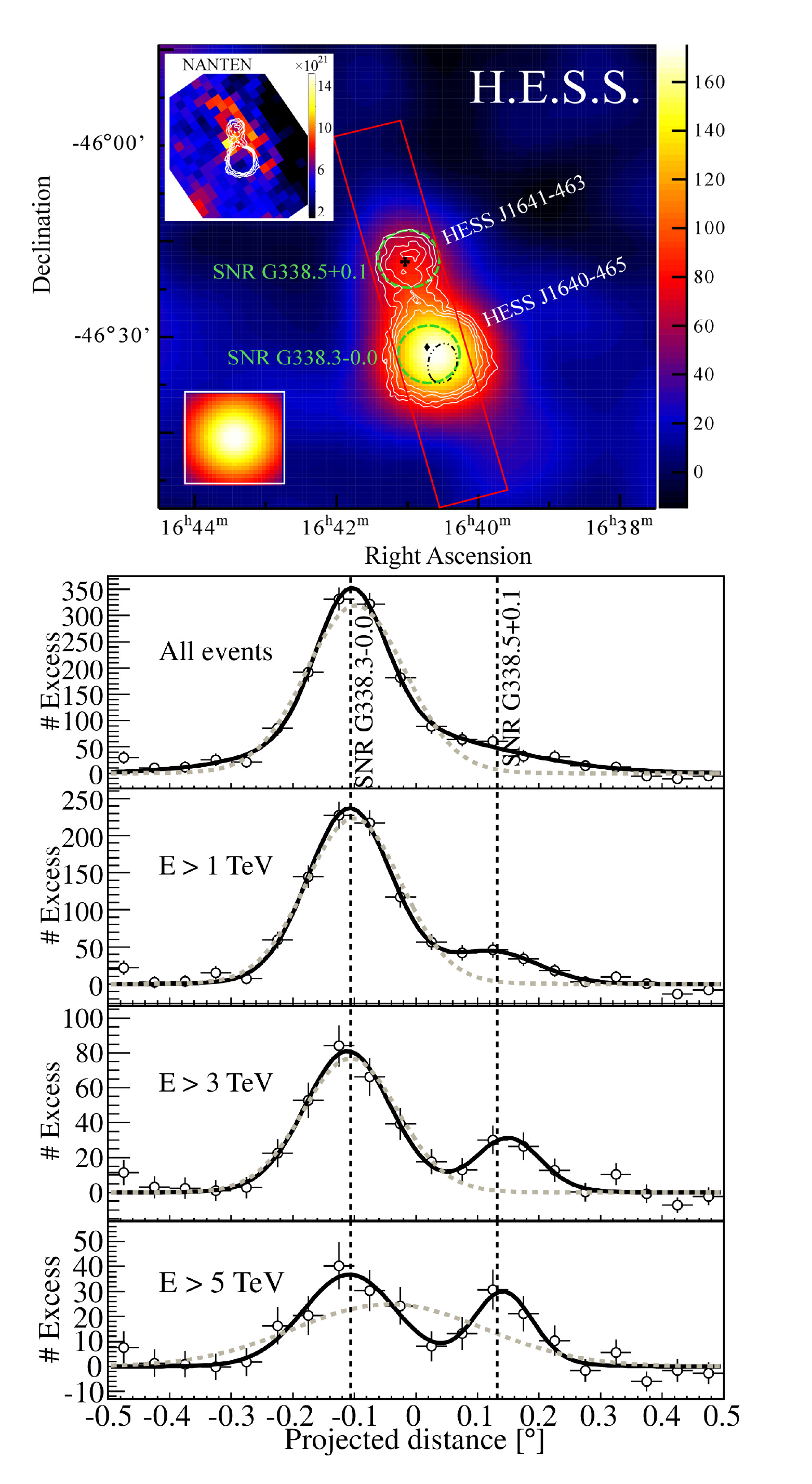} }}%
\qquad
\subfloat[]{{\includegraphics[width=9.75cm]{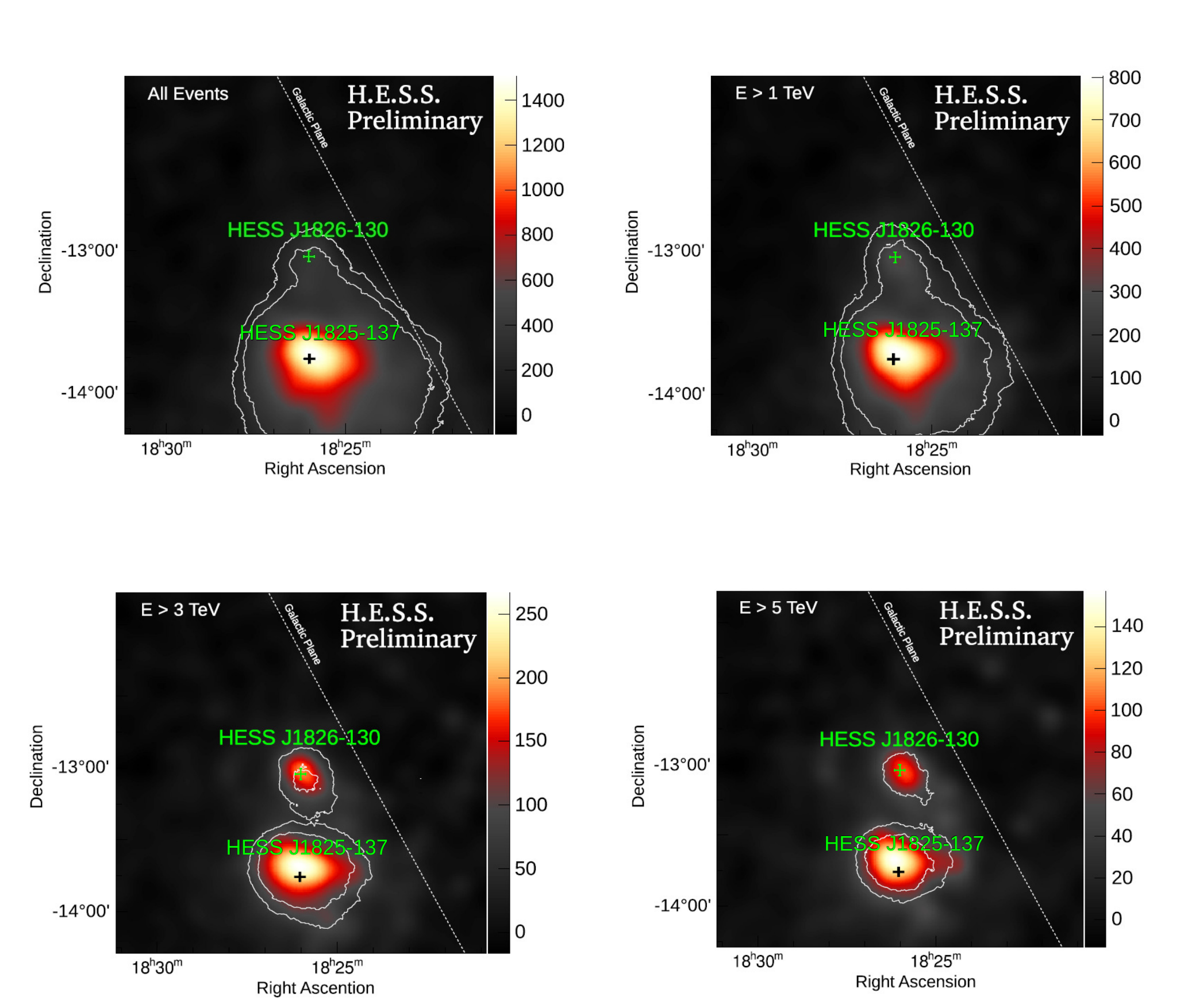} }}%
\caption{(a) Energy dependent excess slices of HESS\,J1641$-$463, the upper panel showing the excess map and the lower panel the distribution of VHE excess events, 
extracted from the red rectangle of the upper panel, and corresponding Gaussian fits in different energy bands. The figures are from \citep{igor}. (b) Energy dependent maps 
of HESS\,J1826$-$130 excess maps for all events and E $>$ 1, 3 and 5 TeV. The white contours indicate the emission regions at 15$\sigma$ and 20$\sigma$ level. The figures 
are from \citep{oziJ1826}.}
\label{sourceConf}
\end{figure}

Note that for such sources suffering from source confusion, an investigation in different energy bands can provide an additional powerful tool for new discoveries \citep{igor}. 
Such an approach has been applied for the cases of HESS\,J1641$-$463 and HESS\,J1826$-$130, since these two sources were hidden under the tails of the bright VHE emission from 
HESS\,J1640$-$465 \citep{J1640} and HESS\,J1825$-$137 \citep{J1825}, respectively. The situation is similar for the case of HESS\,J1741$-$302, but since the nearby bright source, 
HESS\,J1745$-$303 \citep{J1745}, is located farther with respect to the HESS\,J1641$-$463 and HESS\,J1826$-$130 cases, HESS\,J1741$-$302 is less affected by the source confusion. 
Figures \ref{sourceConf}a and Fig. \ref{sourceConf}b show how the visibility of both HESS\,J1641$-$463 and HESS\,J1826$-$130 becomes more prominent as the energy threshold is increased.   

A hard cut configuration was used to optimize the angular resolution, including a minimum of 200 photoelectrons (p.e.) cut for HESS\,J1641$-$463, and 160 p.e for the other 
two sources. The excess maps were fitted by 2D Gaussian models, convolved with the H.E.S.S. point spread function (PSF). The morphological analysis results of these three 
sources are given in Table \ref{morphTable}.

\begin{table*}[ht!]
\scriptsize
\renewcommand{\arraystretch}{1.2}
\centering 
\begin{tabular}{c c c c c c c} 
\hline\hline 
Source Name         & Observation Period &  Live-time & Significance & Best Fit Position                                                           & Extension        & Source Morphology \\ 
                    &                    &  h         & $\sigma$     &      (J2000)                                                                & ($^{\circ}$)     &                   \\
\hline 
HESS\,J1641$-$463   &   2004 - 2011      &  72        & 8.5          & R.A.: 16$^{\textsuperscript{h}}$41$^{\textsuperscript{m}}$2.1$^{\textsuperscript{s}}$    & 0.050            & Point-like         \\
(E $>$ 4.0 TeV)     &                    &            &              & Dec.: $-$46$^{\circ}$18$^{\prime}$13.0$^{\prime\prime}$                     & (UL)             &      \\ \hline
HESS\,J1741$-$302   &   2004 - 2013      &  145       & 7.8          & R.A.: 17$^{\textsuperscript{h}}$41$^{\textsuperscript{m}}$15.8$^{\textsuperscript{s}}$   & 0.077            & Point-like         \\    
(E $>$ 0.4 TeV)     &                    &            &              & Dec.: $-$30$^{\circ}$22$^{\prime}$30.7$^{\prime\prime}$                     & (UL)             &     \\ \hline
HESS\,J1826$-$130   &   2004 - 2015      &  204       & 21.0         & R.A.: 18$^{\textsuperscript{h}}$26$^{\textsuperscript{m}}$0.2$^{\textsuperscript{h}}$    & 0.17 $\pm$ 0.02  & Extended           \\    
(E $>$ 0.5 TeV)     &                    &            &              & Dec.: $-$13$^{\circ}$02$^{\prime}$1.8$^{\prime\prime}$                      &                  &      \\ 
\hline 
 
\end{tabular}
\caption{Morphology analyses results of the sources. The column labelled as observation period shows the year time of the observations, while the live-time column gives the 
corresponding acceptance corrected live-time of the data. The significance columns gives the detection significance of the source. Note that the detection significance and 
morphology model parameters for HESS\,J1641$-$463 are given for the energies above 4.0 TeV. The columns labelled as best fit position and extension give the source 
morphology model parameters obtained from morphology analyses, while the source morphology column indicates whether the source size is compatible with (point-like) or 
significantly larger (extended) than the H.E.S.S. PSF for the corresponding analysis. For the point-like sources, the 99$\%$ upper limits (UL) on the extension are given.} 
\label{morphTable}

\end{table*}

\subsection{Spectral Results}

Source confusion does not only affect source morphologies, but may also potentially distort the observed spectra of VHE $\gamma$-ray sources. This effect is 
negligible ($<$10$\%$) for the case of HESS\,J1741$-$302 when hard cut configuration is used. In the case of HESS\,J1641$-$463, the spectral contamination was 
estimated $\sim$15$\%$ above 0.64 TeV, while it is reduced to $\sim$4$\%$ above 4.0 TeV. Finally, HESS\,J1826$-$130 is the most contaminated VHE source 
in all three sources. The preliminary spectral contamination estimation gives $\sim$40$\%$ below 1.5 TeV and $\sim$20$\%$ above 1.5 TeV. Spectral properties of the 
sources\footnote{HESS\,J1826$-$130 spectral parameters are updated with respect to previously published results \citep{oziJ1826} taking into account the updated 
value of the mirror reflection modelling in H.E.S.S. telescopes.} presented here are given in Table \ref{spectTable} and are affected by contamination. 
The intrinsic (uncontaminated) spectrum of HESS\,J1826$-$130 is expected to be even harder (or having a cut-off at higher energies), given the relatively softer 
spectrum of contaminating emission from HESS\,J1825$-$137. Our preliminary results provide an intrinsic spectrum of HESS\,J1826$-$130 with a spectral index of 
$\Gamma$ = 1.57 $\pm$ 0.15$_{\textsubscript{stat}}$ and a cut-off at 15.2$^{+8.9}_{-4.1}$ TeV.

\begin{table*}[ht!]
\scriptsize
\renewcommand{\arraystretch}{1.2}
\centering 
\begin{tabular}{c c c c c c} 
\hline\hline 
Source Name         & Spectral Model & Normalization (at 1 TeV)                                   & Index                                                     & Cut-off Energy        & Flux ($>$ 1 TeV)   \\ 
                    &                & $10^{-13}$ $\times$ cm$^{-2}$ s$^{-1}$ TeV$^{-1}$          &                                                           & (TeV)                 & (Crab Unit $\%$)            \\
\hline 

HESS\,J1641$-$463   & PL             &  3.91 $\pm$ 0.69$_{\textsubscript{stat}}$ $\pm$ 0.8$_{\textsubscript{sys}}$ & 2.07 $\pm$ 0.11$_{\textsubscript{stat}}$ $\pm$ 0.20$_{\textsubscript{sys}}$ & $-$                   & 1.8                         \\
                    &                &                                                            &                                                           &                       &                             \\ \hline
HESS\,J1741$-$302   & PL             &  2.1 $\pm$ 0.4$_{\textsubscript{stat}}$ $\pm$ 0.4$_{\textsubscript{sys}}$    & 2.3 $\pm$ 0.2$_{\textsubscript{stat}}$ $\pm$ 0.2$_{\textsubscript{sys}}$    & $-$                   & 1.0                         \\    
                    &                &                                                            &                                                           &                       &                             \\ \hline
HESS\,J1826$-$130   & ECPL           &  8.28 $\pm$ 0.68$_{\textsubscript{stat}}$ $\pm$ 1.6$_{\textsubscript{sys}}$ & 1.66 $\pm$ 0.11$_{\textsubscript{stat}}$ $\pm$ 0.20$_{\textsubscript{sys}}$ & 13.5$^{+4.7}_{-2.7}$  & 4.0                         \\    
                    &                &                                                            &                                                           &                       &                             \\ 
                    
\hline 

\end{tabular}
\caption{Spectral parameters of the sources presented in this work. The column labelled as spectral model indicates whether the preferred spectral model of the source is a 
simple power-law function (PL) or a power-law with an exponential cut-off (ECPL) function. The normalization, index and cut-off energy columns give the corresponding model 
parameters, while the flux column gives the corresponding integrated flux above 1 TeV in the units of Crab Nebula flux above same energy.}  

\label{spectTable}

\end{table*}

\begin{figure}[ht!]
\centering
\subfloat[]{{\includegraphics[width=7cm]{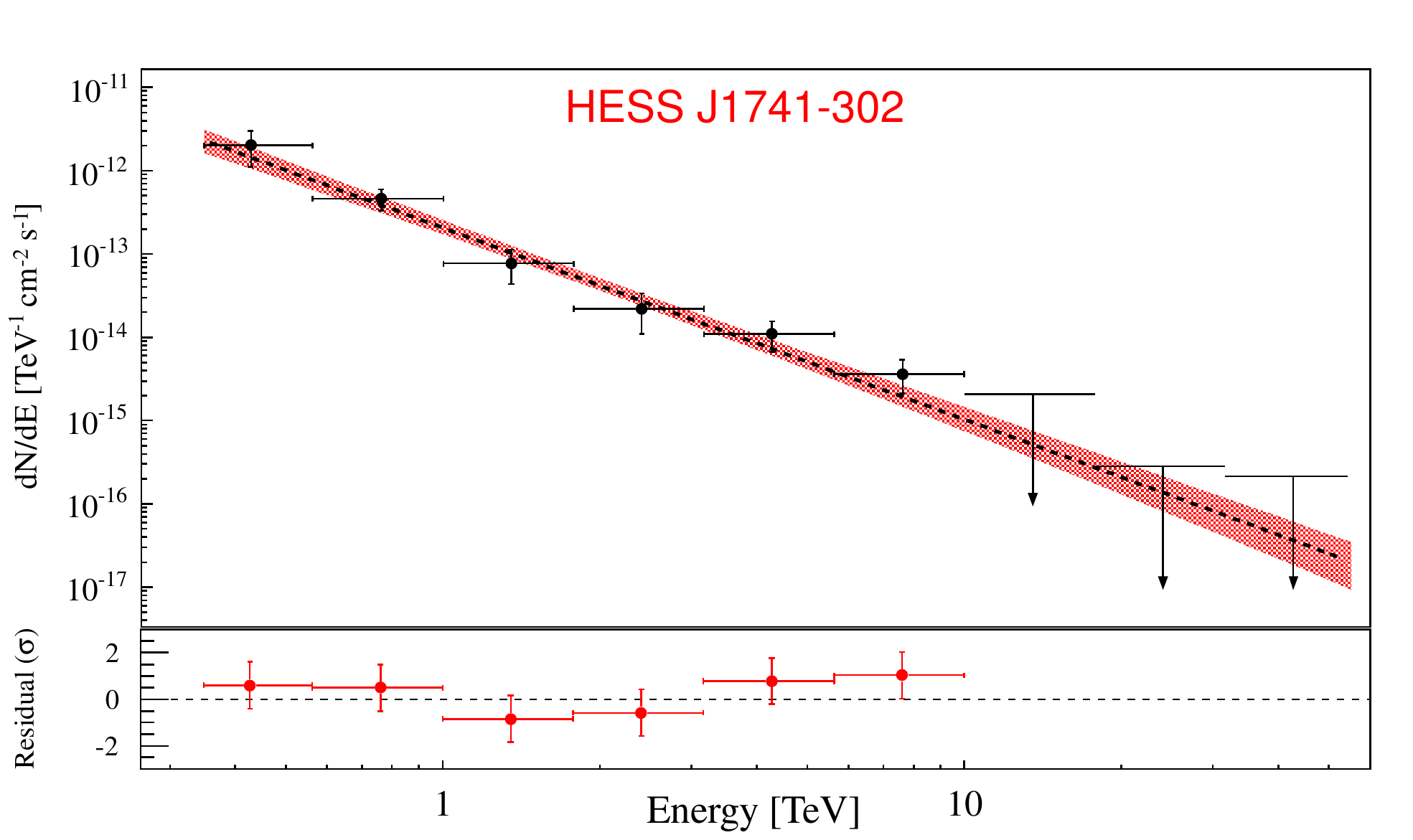} }}%
\qquad
\subfloat[]{{\includegraphics[width=7cm]{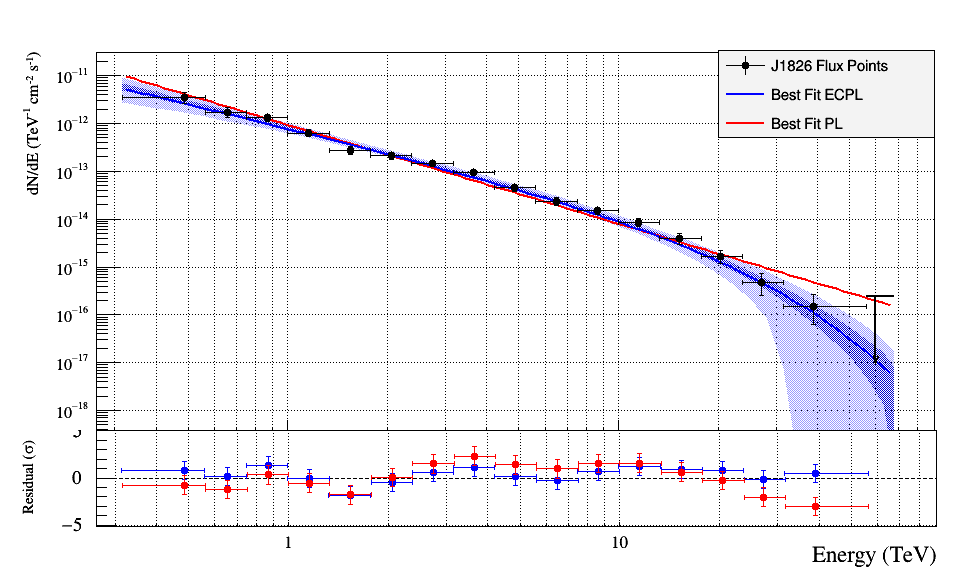} }}%
\qquad
\caption{VHE $\gamma$-ray spectra of HESS\,J1741$-$302 \citep{oziJ1741} (a) and HESS\,J1826$-$130 \citep{oziJ1826} (b). The red and blue shaded areas represent the 68$\%$ 
confidence interval for the fitted spectral model of PL (a) and ECPL (b), while the light blue shaded region represents the 99$\%$ confidence interval of the ECPL model.}%
\label{spectra}
\end{figure}

\subsection{Interstellar Medium}

A characteristic shared among these three sources is that there are dense gas regions coincident with their best fit positions. These dense gas regions can provide rich target material 
for accelerated particles to produce VHE $\gamma$-ray emission via proton-proton interactions followed by a subsequent $\pi^{0}$ decay \citep{aha90}, especially in case a CR accelerator 
(such as a SNR) is located close to, or coincident with, the sources of interest. For investigating a possible hadronic origin of the emission detected from these sources, the analysis of 
data from available surveys of atomic and molecular hydrogen around the location of the sources have been carried out. The distribution of molecular gas around the sources presented in 
this work is obtained by integrating the $^{12}$CO 1$\rightarrow$0 rotational line emission measured with the NANTEN Sub-millimeter Observatory \citep{nanten}. 

The distribution of molecular gas around HESS\,J1641$-$463 is obtained by integrating over a range in velocity between $-$40 km s$^{-1}$ to $-$30 km s$^{-1}$. The total 
column density from the extraction region of the source is 1.7 $\times$ 10$^{22}$ cm$^{-2}$, while the density and the total mass are about 100 cm$^{-3}$ and 2.4 $\times$ 
10$^{5}$ solar masses at 11 kpc, respectively \citep{igor}. In the case of HESS\,J1741$-$302, eight molecular clouds (MCs) found coincident with the best fit position of the source along the 
line of sight. Integration velocity intervals change between $+$7 km s$^{-1}$ and $-$221 km s$^{-1}$ depending on the corresponding molecular cloud (MC). Analysis of these MCs give 
the densities and total masses changing in the range of [62, 380] cm$^{-3}$ and [1.9, 9.8] $\times$ 10$^{5}$ solar masses at the corresponding MC distances, respectively (see 
\citep{oziJ1741} for the details of the interstellar medium study). The analysis of interstellar medium around HESS\,J1826$-$130 points out two giant molecular clouds (GMC) with few 
10$^{5}$ solar masses at 3.7 kpc and 4.7 kpc \citep{voisin}. The ambient gas densities of these GMCs are estimated of the order of 600 cm$^{-3}$.

\section{DISCUSSIONS}

Investigation of VHE $\gamma$-ray sources showing hard spectrum without a clear cut-off is important to understand the origin of the highest energy CRs close to the knee in our Milky Way 
Galaxy, especially in the presence of dense gas regions coincident with the sources. The sources presented in this work show common characteristics in this sense, thus indicating possible 
hadronic VHE emission origin. Assuming that the VHE emission originates from hadronic processes happening in the dense gas regions \citep{kelner} and taking into account the properties of 
these regions coincident with these sources, one can conclude that the spectra of parental particle population can extend up to at least several hundreds of TeV. The total energy required 
in protons, $W_{pp}$, to produce the inferred $\gamma$-ray luminosity, $L_{\gamma}$ = 4$\pi$ $F_{\gamma}$($>$ 0.4 TeV) $D^{2}$, can be estimated as $W_{pp} = L_{\gamma}$ $\times$ $t_{pp}$, 
where $t_{pp}$ = 5.76 $\times$ 10$^{15}$ $\times$ ($n_{gas}$/cm$^{-3}$)$^{-1}$ s is the cooling time for proton-proton collisions. Calculations of the energy budget for these sources gives 
that $W_{pp}$ = $\sim$10$^{48}$ erg and $W_{pp}$ = $\sim$10$^{47}$ erg for HESS\,J1641$-$463 \citep{igor} and HESS\,J1826$-$130 \citep{oziJ1826}, for the distances of 11 kpc and 4 kpc, 
respectively. In the case of HESS\,J1741$-$302, $W_{pp}$ is found between 7.0 $\times$ $10^{46}$ erg and 1.5 $\times$ $10^{48}$ erg depending on gas densities calculated for each MC along 
the line of sight \citep{oziJ1741}.

In any case, our studies do not allow us to distinguish whether the VHE $\gamma$-ray emission from these sources has leptonic or hadronic origin. Taking into account the point-like 
morphologies of HESS\,J1641$-$463 and HESS\,J1741$-$302, binary scenarios can also be envisaged. The fact that no variability could be observed from these sources can not be taken 
as evidence for disfavoring binary origin because of the low statistics, while the lack of an optical counterpart can be related to high optical extinction or the location 
of sources close to the Galactic plane. One can not exclude a cut-off in the $\gamma$-ray spectra of these two sources because of the limited statistics above 10 TeV. In addition, 
a leptonic scenario, where electrons accelerated by the pulsar PSR\,J1826$-$1256 are up-scattering CMB or IR photons, can also explain the VHE emission from HESS\,J1826$-$130. Such as hard 
spectrum at H.E.S.S. energies can be produced by an uncooled electron population with spectral index close to 2.0 and a cut-off at around 70 TeV. The energy output in accelerated electrons 
is 2 $\times$ 10$^{47}$ ergs for a distance of 7 kpc. This source has a spectrum very similar to other PWNe, in particular, Vela X \citep{vela}. HESS\,J1826$-$130 could be an indication 
of a distinctive PWN population, with very hard spectra and relatively high cut-off energies

\section{CONCLUSIONS}

The analysis of the three sources presented here shows the common characteristic of very hard spectra and plausible scenarios where parental population of hadrons extend up to several 
hundreds of TeV, suggesting they may be representing a population of CR accelerators active in the Galaxy. However, an interpretation based on leptonic scenarios can not be discarded.

The future Cherenkov Telescope Array (CTA), with its much better angular and energy resolution and sensitivity, with respect to the current imaging air Cherenkov telescope systems, will 
facilitate a deeper study and better constrain the spectral parameters of these sources . In addition, CTA will be able to detect many more VHE $\gamma$-ray sources throughout the Galaxy, 
some of them perhaps sharing characteristics with the sources presented here. In conclusion, the enhanced capabilities of CTA will allow us to establish if such a population of CR accelerators 
active in the Galaxy.

\section{ACKNOWLEDGEMENTS}
The support of the Namibian authorities and of the University of Namibia in facilitating the construction and operation of H.E.S.S. is gratefully acknowledged, as is the support by the German 
Ministry for Education and Research (BMBF), the Max Planck Society, the German Research Foundation (DFG), the French Ministry for Research, the CNRS-IN2P3 and the Astroparticle 
Interdisciplinary Programme of the CNRS, the U.K. Science and Technology Facilities Council (STFC), the IPNP of the Charles University, the Czech Science Foundation, the Polish Ministry of 
Science and Higher Education, the South African Department of Science and Technology and National Research Foundation, the University of Namibia, the Innsbruck University, the Austrian Science 
Fund (FWF), and the Austrian Federal Ministry for Science, Research and Economy, and by the University of Adelaide and the Australian Research Council. We appreciate the excellent work of the 
technical support staff in Berlin, Durham, Hamburg, Heidelberg, Palaiseau, Paris, Saclay, and in Namibia in the construction and operation of the equipment. This work benefited from services 
provided by the H.E.S.S. Virtual Organisation, supported by the national resource providers of the EGI Federation. This research has made use of software provided by the {\it Chandra} X-ray 
Center (CXC) in the application packages CIAO, ChIPS, and Sherpa. This research has made use of the SIMBAD database, operated at CDS, Strasbourg, France. This research has made use of the 
ATNF pulsar catalog database (http://www.atnf.csiro.au/research/pulsar/psrcat/). The NANTEN project is based on the mutual agreement between Nagoya University and the Carnegie Institution of 
Washington. Sabrina Casanova and Ekrem O\u{g}uzhan Ang\"uner acknowledge the support from the Polish National Science Center under the Opus Grant UMO-2014/13/B/ST9/00945.

\end{document}